*Note to reader: This chapter was primarily written in 2011 and will be published in a book,(due to be published in 2016). This can be cited according to its arXiv reference. This is the second version I have submitted to arXiv, with relatively minor changes mostly to wording.*

# Predicting distributions of invasive species

Jane Elith

School of BioSciences, The University of Melbourne 3010. Australia

j.elith@unimelb.edu.au

## Abstract

This chapter aims to inform a practitioner about current methods for predicting potential distributions of invasive species. It mostly addresses single species models, covering the conceptual bases, touching on mechanistic models, and then focusing on methods using species distribution records and environmental data to predict distributions. The commentary in this last section is oriented towards key issues that arise in fitting, and predicting with, these models (which include CLIMEX, MaxEnt and other regression methods). In other words, it is more about the process of thinking about the data and the modelling problem (which is a challenging one) than it is about one technique versus another. The discussion helps clarify the necessary steps and expertise for predicting distributions. Some researchers are optimistic that correlative models will predict with high precision; while that may be true for some species at some scales of evaluation, I believe that the issues discussed in this chapter show that substantial errors are reasonably likely. I am hopeful that ongoing developments will produce models better suited to the task and tools to help practitioners to better understand predictions and their uncertainties.

## 5.1    Introduction

In a newly invaded region, invasive species can progress through the stages of introduction, establishment and dispersal to a full range. There is currently much worldwide interest in predicting distributions of invasive species, and many organisations will be faced with questions of whether and how to embark on such a task, or how to interpret predictions that others have provided. This chapter provides information on predicting the final stage, commonly referred to as the potential distribution, of the species in the invaded range. In contrast, Chapter 6 discusses methods for modelling the whole invasion process.

The names for these predictions of invasive species distributions can be confusing because the same terms can be used for distinctly different aims and models. So here, regardless of other uses of the words, mention of pest risk mapping, climate matching, niche mapping and predicting potential distributions will all mean the same thing: a model or process that aims to produce a map of areas that are likely to be suitable for the species. The advantages of these maps are obvious: species can be screened for those likely to become pests (i.e. likely to cause harm), monitoring programs can target areas most likely to be infested, arrangements can be established for cost-sharing between jurisdictions over a large region and so on (Brunel *et al*., 2010; Cook *et al*., 2007; Richardson & Thuiller, 2007).

Many governments, agencies and organisations now invest in some form of pest risk mapping. As yet, there appears to be no complete system for mapping; most are examples, or case studies for particular species, or prototype systems. For instance, Pratique (https://secure.fera.defra.gov.uk/pratique/index.cfm) is a European Union initiative broadly targeting pest risk analysis, but with components focusing on mapping ranges. In the United States, the Animal and Plant Health Inspection Service conducts risk assessments using NAPPFAST (Magarey *et al*., 2007), while in Australia, the Department of Agriculture, Forestry and Fisheries uses a simple climate matching system (CLIMATE) to predict climate suitability for species of biosecurity concern (e.g. Bomford *et al*., 2007). Globally, there is interest in linking biodiversity databases with modelling tools to facilitate pest risk mapping anywhere in the world

(http://wiki.tdwg.org/InvasiveSpecies), but there is understandable uncertainty about the likely quality of the outputs.

This chapter begins with a brief discussion of approaches for modelling broad ecological units or climates (Section 5.2). The focus then shifts to single species models, covering the conceptual bases (Section 5.3), touching on mechanistic models (Section 5.3), and then focusing on methods using species distribution records and environmental data to predict distributions (Section 5.5). The chapter includes a mix of commentary based on my own research, review and advice, with the intention of providing interpretation of the current state of the science and commentary on useful ways forward.

## 5.2    Community or climate-based mapping

Some approaches to modelling potential ranges of invasive species focus on biological or environmental units aggregated above the species level. For instance, Richardson and Thuiller (2007) predicted the global distribution of seven South African biomes. They suggested that the results, which were essentially a biologically oriented climate matching, would be useful for screening species' introduction risks. Baker *et al.* (2000) reviewed applications of climate-based mapping that mapped climate without reference to species responses, giving examples both in environmental space (e.g. the early climographs of Cook, 1925) and geographic space (e.g. the Match Climates option in CLIMEX; see Box 5.2 and Sutherst, 2003). Brunel *et al.* (2010) proposed that Köppen-Geiger climate zones and world hardiness zones provide ecoclimatic information relevant to screening potential invasive plant species for the European and Mediterranean Plant Protection Organization. Thomas and Ohlemüeller (2010) used rainfall and temperature information to map similar climates both locally (within 1000km of a target cell) and globally. They then estimated likelihood of invasion (*invasibility*) by assuming that similar non-local climates represent potential source locations of invasive species. Their maps comparing risks under current and future climates suggested increases in invasibility with climate change (e.g. Figure 5.1).

<Figure 5.1 here>  (this figure will be from Thomas & Ohlemüeller, 2010, their Figure 2.2B).

These types of models or data summaries can be used to develop an understanding of general patterns of invasions. They can also give a broad overview of whether a region is even remotely likely to be suitable for a species of concern (or alternatively, whether the climates of two regions overlap and, therefore, whether one poses a potential risk for the other). In that sense, these models could be considered useful background information or a first step for assessing invasive potential.

## 5.3    The conceptual basis for predicting potential distributions of invasive species

In many situations, predictions are needed for a particular species. Users require mapped estimates of where species could persist in a given region, and this is related to questions about the biotope – i.e. the geographic location of the species' niche. In the species modelling arena most niche definitions rely on Hutchinson's viewpoint (Hutchinson, 1957) – namely that the fundamental niche is a multi-dimensional hyper-volume with 'permissive conditions and requisite resources as its axes' (Colwell & Rangel, 2009, p. 19651), in which every point corresponds to a state that would allow the species to exist indefinitely. The dimensions of this niche are limited to the subset of all possible conditions that directly affect the fitness of the organism (Kearney, 2006). In practice, modellers often focus on the species' response to climate, although this is neither essential nor most relevant for some species and spatial extents (Hulme, 2003). For a clear explanation of Hutchinson's niche ideas, the links between niche (environmental) and biotope (geographic) space, and implications for species modelling, see Colwell and Rangel (2009).

The full fundamental niche need not be apparent at a given time. The concept of the *potential* niche was introduced by Jackson and Overpeck (2000) to describe those portions of the fundamental niche (those environments) that actually exist somewhere in geographic space at a specified time. The idea of modelling the potential distribution of an invasive species in a region is related to this definition. The *realised* niche (where the species actually occurs) is usually a smaller environmental volume (or geographic area) than the fundamental and potential niches. Hutchinson (1957) saw the realised niche as a subset of the fundamental niche, limited by biotic interactions – for instance, by the presence of competitors or predators, or the absence of mutualists. Others (e.g. Pulliam, 2000) refined the definition to allow for source-sink theory and dispersal limitations. Thus, sink populations can allow the realised niche to be larger than the fundamental niche, and constraints to dispersal and past disturbances can limit the realised niche beyond the effects of biotic interactions.

These differences between the realised and fundamental niches are relevant to invasive species, particularly when we consider the realised niche in native ranges versus the global potential or fundamental niche. Invasive species often persist in environments in their invaded ranges that either were not occupied by them (because of dispersal or biotic limitations) or were non-existent in their native range. That is, invasive species are able to expand into parts of their fundamental niche that are not available in their native range (Le Maitre *et al*., 2008). Methods best suited to modelling the potential distribution of an invasive species in any new region are therefore those that most directly estimate the fundamental niche. While these will usually overestimate the final distribution of the invasive species in the invaded range, they will at least show what areas could be occupied if the species is able to spread everywhere and if biotic conditions are suitable.

A final complication in modelling invasive species is that their spread may not simply represent the expression of the fundamental niche as set by the gene pool in their native range. Instead, new conditions in the invaded range may provoke adaptive evolution (Colwell & Rangel, 2009; Huey *et al*., 2005). While not a priority for this chapter, methods for exploring adaptive genetic change and predicting traits likely to be under selection pressure are relevant to invasive species and are an important topic for understanding the ecology and biogeography of invasive species (Ackerly, 2003; Alexander & Edwards, 2010).

## 5.4     Methods aiming to model and map the fundamental niche: mechanistic models

Section 5.3 provides reasoning for preferring methods that model biological traits that are directly related to the fundamental niche of the species. I refer to these as *mechanistic models* because they focus on mechanisms or processes rather than patterns. Mechanistic models could – depending on the way the model is set up – include eco-physiological models, biophysical models, life-history models, phenological models, foraging energetic models and models based on functional traits (Buckley *et al*., 2010; Kearney & Porter, 2009; Morin & Lechowicz, 2008). For our purposes, the main criterion for considering a model to be mechanistic is that it attempts to capture the dominant processes contributing to survival and fecundity, and it links these processes to environmental data in a way that enables mapped predictions of the niche. These models are not fitted to species location data, and are, therefore, free from the problem that occurrence records are tied to the realised niche. Instead, they focus on the processes and physiological limits that constrain the distribution and abundance of a species.

Kearney and Porter (2009) review the potential to apply principles of biophysical ecology to modelling species distributions and include information on how to model key functional traits of a range of organisms (e.g. dry-skinned and wet-skinned ectotherms, endotherms, aquatic organisms and plants). Their software (NicheMapper; http://www.zoology.wisc.edu/faculty/por/por.html) is available, although it is quite complex to use and further development is underway to make it more broadly accessible (M. Kearney, personal communication, 2014). Examples of applications include Kearney and Porter (2004), Kearney *et al*. (2008, 2010) and Porter *et al*. (2002). These models require information on the morphology, physiology and behaviour of species (e.g. how endotherms balance metabolic rate and heat loss at various temperatures), and a means for translating the environment experienced by the animal to the landscape-scale geographic information system data usually available for mapping.

In related examples, Buckley *et al*. (2010) use three mechanistic models (a biophysical model, a life-history model and a foraging energetic model) to model a butterfly and a lizard; Morin and others (Chuine &

Beaubien, 2001; Morin & Lechowicz, 2008; Morin & Thuiller, 2009) use a phenological model, Phenofit, to model trees. Phenofit focuses on the impacts of physiological stress on fitness, and on the synchronisation of developmental stages with seasonal variations in climate (Morin & Thuiller, 2009).

These authors and others (e.g. Hijmans & Graham, 2006) have compared mechanistic models with *correlative* models based on relationships between observed species locations and measured or estimated environmental conditions. These comparisons often show congruence of predictions in the regions in which the correlative model was trained, and a range of outcomes (from congruence to dissimilarity) for predictions for novel times or places (Kearney *et al*. 2010; Morin & Thuiller 2009). Kearney and Porter (2009) compare the likely strengths and weaknesses of mechanistic and correlative models, and Dormann *et al*. (2012) provide an interesting discussion of the apparent dichotomy between mechanistic and correlative models.

Mechanistic models are the subject of active research programs, but are less frequently attempted than correlative models owing to the complexity of the models and the time it takes to gather appropriate data and fit models. It is conceptually appealing to focus on process and understand the constraints to distribution, because these will then be applicable to any geographic region or future time, providing the species does not evolve different tolerances in new environments. Despite the fact that mechanistic models are theoretically well suited to invasive species and several reviews recommend them (e.g. Buckley *et al*., 2010; Gallien *et al*., 2010; Kearney & Porter, 2009), few applications to invasive species exist (but see Elith *et al*. (2010) and Kearney *et al*. (2008) for a cane toad example). Of course, even though compatible with the modelling problem, mechanistic models will not be perfect. The most likely errors and uncertainties stem from the need to identify key processes (is there enough information to pinpoint these, and is the model sufficient to include and combine them appropriately?), parameterise the models appropriately (are relevant experimental data available? Buckley *et al*., 2010; Kearney & Porter, 2009), and match microclimate or laboratory measurements to the broad scale climatic variables available for mapping. Given the time and expertise needed to fit mechanistic models, I expect them to be most useful for species of exceptional importance, or as a guide to likely distributions if generalised versions can be made available to serve as templates for sets of physiologically similar species.

## 5.5    Methods that use information on the realised niche

Most predictions of a species' invasion potential are based on models fitted to observed location data (Venette *et al*., 2010). Data from the native range (and perhaps additional records) are used to characterise and predict suitable conditions elsewhere. The commentary in this section is oriented towards key issues that arise in fitting, and predicting with, correlative models. In other words, it is more about the process of thinking about the data and the modelling problem than it is about one technique versus another. This reflects my viewpoint that the issues are critically important, and the modelling problem is one that requires careful thought.

Throughout, I will use the term *correlative models* (see Box 5.1 and Dormann *et al*., 2012) to refer to most of these models because they are pattern-based models that quantify the relationship between a species presence (or presence-absence or abundance) and a set of environmental covariates. That is, I use *correlation* in the broad sense of relationships between variables, in this case between a response (the species) and one or more predictors or covariates. A model that does not fall completely into this class is CLIMEX (Box 5.2), which relies on species records but has a more process-based orientation than correlative species distribution models (SDMs). The term *pest risk models* will include CLIMEX, but SDMs or correlative models will not. This is for convenience of discussion; obviously CLIMEX could also be termed a SDM. Box 5.1 provides background to the more general (and original) use of correlative models for modelling species other than invasive species and introduces the phrase *equilibrium SDM* for such applications, Box 5.2 describes CLIMEX, and Box 5.3 outlines the broad classes of correlative models. Table 5.1 summarises key references and examples of invasive species applications. If you are unfamiliar with correlative models, reading the Boxes should give enough background for the following sections. Note that correlative models – sometimes with additional components to include processes of dispersal – have been used to fit and predict distributions entirely in the invaded range. These models are generally not considered here (but see Section 5.5.2) because they require specialised methods and are usually only relevant where a species has been in a country for a considerable time.

## Box 5.1 The general use of correlative models in ecology

Correlative methods include a range of techniques variously referred to as species distribution models (SDMs), ecological niche models, bioclimatic envelopes, profile methods or climate matching techniques. None of these were originally designed to model invasive species. Instead, they were intended for modelling (and perhaps mapping) a species–environment relationship, but only using the current distribution of the species within the sampled geographic extent (Elith & Leathwick, 2009b). I will refer to this original use as *equilibrium SDM*, even though ecologists will recognise that use of the word *equilibrium* opens up many questions about time frame, dispersal barriers, effects of disturbance and so on (Franklin, 2010; Peterson *et al*., 2011). It is important to keep this history in mind when reading the SDM literature and when considering the range of methods available because the history provides context for interpreting what people have done and why they have done it. For instance, some equilibrium SDMs use geographic space rather than environmental space as the predictors of occurrence (e.g. convex hulls, kernel density estimators and kriging; Elith & Leathwick, 2009b). These might be useful where data are very sparse or where geographic space strongly determines distributions, but they are not useful for predicting the distribution of invasive species in new geographically remote areas. The more common use of environmental predictors is based on the belief that – at most scales and in most regions – environment is important in structuring distributions (Section 5.5.4).

The literature on SDMs has expanded rapidly in the last 10 years, and tutorials, books and reviews are regularly emerging; see, for example, Austin (2002, 2007), Elith and Leathwick (2009b), Franklin (2010), Guisan and Thuiller (2005), Guisan and Zimmermann (2000); Pearson (2007), Peterson *et al*. (2011), and Schröder (2008). Equilibrium SDMs have been fitted for terrestrial, marine and freshwater species, and from macroecological (coarse grain, large extent) to local (fine grain, small extent) scales. Models using well-designed survey data and ecologically relevant predictor variables have produced useful insights and reliable predictions to new sites within the sampled regions (Bio *et al*., 2002; Leathwick & Austin, 2001; Ysebaert *et al*., 2002). Predictions have provided key inputs for conservation planning and resource management, identifying new sites for rare species surveys, and global analyses of species distributions (Ferrier, 2002; Fleishman *et al*., 2001; Rangel *et al*., 2006; Zimmermann *et al*., 2007). Because equilibrium SDMs aim to predict within the range of the training data, users have tended to evaluate their performance at points within that range (e.g. using cross-validation) or by assessing whether the modelled relationships are ecologically sensible.

Table 5.1 Example correlative methods for modelling species distributions

| General class | Model (abbreviation) | Species data | Partial plots for effect on response | Comment | References for (a) explaining model and (b) invasive application |
|---|---|---|---|---|---|
| Expert model | Habitat suitability index (HSI) | Expert | Yes | Use expert knowledge for shape of species response | (a) Burgman *et al.* (2001) (b) Inglis *et al.* (2006) |
| Expert model | Expert | Expert/presence | No | Use expert knowledge to select variables and perhaps to inform about presence | (a, b) Rodda *et al.* (2009) |
| Climate envelope | BIOCLIM | Presence | No | Delimits climate envelope only using presence data, sometimes using percentiles; prediction from most extreme (limiting) variable | (a) Busby (1991) (b) Booth (1988) |
| Machine learning | One-class support vector machines | Presence | No | Few uses, but being included in some ensembles | (a) Hastie *et al.* (2009) (b) Guo *et al.* (2005); Drake & Bossenbroek (2009) |
| Factor analysis | Ecological niche factor analysis (ENFA) | Presence-background | No | Also known as Biomapper | (a) Hirzel *et al.* (2002) (b) Steiner *et al.* (2008) |
| Machine learning | Genetic algorithm for ruleset production (GARP) | Presence-background | No | Widely used; final model is an average over best selected rules | (a, b) Peterson (2003) |
| Machine Learning | Maximum entropy (MaxEnt) | Presence-background | Yes | Widely used; complexity of model can be adjusted by choice of features and adjusting regularisation | (a) Phillips *et al.* (2006); Elith *et al.* 2011 (b) Rodda *et al.* 2011 |
| Regression | Generalised linear models (GLMs) or generalised additive models (GAMs) | Various | Yes | Statistical regression methods; generalised additive models allow smoothed data-driven functions | (a) Hastie *et al.* (2009) (b) Mellert *et al.* 2011 |
| Regression | Non-parametric multiplicative regression | Various | Yes | Implemented in Hyperniche; only found invasive examples use invaded range data | (a) McCune (2006) (b) Reusser & Lee (2008) |

| Machine Learning | Decision tree | Various | Yes | Also known as classification and regression trees; more often used for decision analysis (e.g. on whether species will become invasive or not) | (a) Hastie *et al*. (2009); De'ath & Fabricius (2000)<br>(b) Václavík & Meentemeyer 2009 (only in invasive range) |
|---|---|---|---|---|---|
| Machine Learning | Ensembles of trees: boosted regression trees (BRT), or random forests (RF) | Various | Some | Most invasive species examples are within ensembles; automatically model interactions unless stumps used | (a) Hastie *et al*. (2009)<br>(b) Broennimann *et al*. 2007 |
| Machine Learning | Artificial neural nets | Various | Some | One of the earliest machine learning methods to be used in species modelling; regarded as a good general purpose algorithm | (a) Hastie *et al*. (2009)<br>(b) Gevrey and Worner (2006) |
| Ensembles | Ensembles of any type of models | Not applicable | No | Several examples emerging, with varied approaches for selecting the component models | (a) Thuiller (2003)<br>(b) Broenniman *et al*. (2007); Stohlgren *et al*. (2010) |

### 5.5.1 Issue 1: What niche can be characterised by these models?

Section 5.3 discusses fundamental and realised niches, a critical issue for pest risk models. The dual concepts of environmental (niche) and geographic (biotope) space make it clear that to characterise the environmental niche well, records of species locations must be taken from regions in which the species has had opportunity to spread (geographically) to all suitable locations. Hence, it is logical to focus on places where the species is most likely to be at equilibrium (i.e. the native range).

It is not possible to make a definitive statement about exactly what niche is being modelled by equilibrium SDMs (Box 5.1), but it is most closely related to the realised niche (Austin, 2002; Austin *et al.*, 1990; Colwell & Rangel, 2009; Jiménez-Valverde *et al.*, 2008; Soberon & Nakamura, 2009). The species data, choice of predictor variables and modelling method all affect the outcome. For instance, imagine being fortunate enough to have a large, comprehensive and unbiased sample of the abundance of a species across its whole range. From these data, one might expect to successfully model the realised niche. However, if the available predictor variables lack some important dimension of the niche (e.g. soil phosphorus for plants needing high levels of phosphorus) or the modelling method is incapable of fitting the shape of the true relationship, then the niche will be imperfectly modelled. The aim, therefore, in fitting an SDM for an invasive species is to do as much as possible to characterise the realised niche well (excluding sink populations), and beyond that, to move towards approximating the fundamental niche. An early application of this idea (Booth *et al.*, 1988) expanded the native range climatic profile for 13 eucalypt species using forestry trial plot results from Africa, intending to better characterise the fundamental niche to inform successful tree introductions for plantations. Sections 5.5.2 to 5.5.7 include discussion on how species records, predictors, the model and the prediction extent all affect how accurately the realised niche is modelled, and resulting implications for prediction of invasive potential.

Similar issues apply to CLIMEX (Box 5.2) because the model is often primarily fitted using location data. The CLIMEX predicted distribution may be closer to the realised niche than the fundamental niche, depending on the extent to which the dispersal of the species has been limited and on the amount of additional physiological data (Lawson *et al.*, 2010). Physiological data, if reliable and if successfully rescaled to be consistent with the predictor information, should allow the prediction to edge closer to the fundamental niche (Box 5.2).

For predicting potential distributions of invasive species, one drawback of being tied to observation records is that biotic interactions affect the outcome: the realised niche in the native range is usually affected by pathogens, pests, competitors and predators. In some instances, invasive species have shown evidence of release from inhibiting biotic factors, and models from the native ranges where biotic interactions were important but unquantified have not been good predictors of distributions in the invaded range (Le Maitre *et al.*, 2008). This is an inherent weakness of models based on the realised niche. Biotic interactions are notoriously difficult to include as predictors because their effects are almost always confounded with the effects of other covariates (Leathwick & Austin, 2001). Researchers often assume that biotic interactions vary enough across the species range so that a reasonably sized sample will smooth over local biotic effects. This will only sometimes apply, and the use of these models for predicting other than the realised niche is problematic. Solutions may not exist, but one way to counteract this problem is to collate available knowledge on the impact of biotic interactions on the native range of a species and use that as a guide to likely errors in predicted distributions. Further, recent progress in methods for modelling species co-occurrences (Ovaskainen *et al.*, 2010; Pollock *et al.*, 2014) can provide strong inference about likely inter-species effects. However, SDMs for species without significant pathogens, pests and competitors are likely to be the most accurate.

## Box 5.2 CLIMEX

CLIMEX is a commercially available modelling method that was first published in the 1980s and has now been applied to many species and adopted worldwide in various agencies and governmental departments (Sutherst, 2003; Sutherst & Maywald, 1985). It was specifically developed for modelling invasive species. The primary output is a mapped prediction of the favourability of a set or grid of locations for a given species. The model also produces a suite of information to allow further understanding of species response to climate. CLIMEX requires location records of a species in its native range, and uses these with climate data and other optional information (locations of persistent populations in invaded regions, relative abundance, seasonal phenology and laboratory data) to infer a species' climatic requirements. The model is based on population process concepts of how a species responds to environment, and attempts to characterise growth and stress responses to weekly climatic conditions. The current version (version 3; Sutherst *et al.*, 2007) of CLIMEX includes six growth indices (temperature, moisture, light, radiation, substrate and diapause/dormancy) over which a seventh index, biotic interactions, can be used as a multiplier. There are up to eight stress indices based on temperature and moisture (heat, cold, dry, wet, and their interactions, e.g. hot and dry) plus two constraints to persistence that can be imposed over all others: length of growing season and obligate diapause/dormancy/vernalisation. The indices and constraints aim to cover the major mechanisms by which terrestrial species respond to their environments.

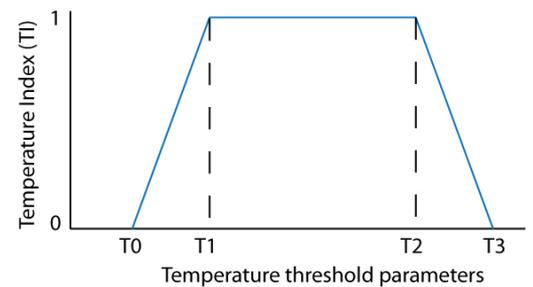

The model is conceptualised as providing two main seasons for the species: one for population growth and one for population survival. This is directly relevant to invasive species because new geographical regions can be determined as holding suitable environments for population persistence or population growth, the latter most related to pest status. In fitting the model, decisions are required about which indices or constraints are relevant to the species, and how to estimate their parameters. Growth indices relate to seasonal population growth and mostly require four parameters to be set (see inset graph in which parameters are T0 to T3). Stress indices are defined by a threshold value and an accumulation rate, and stress is assumed to accumulate exponentially with time. Parameters are often set by starting with template values and then iteratively altering them and assessing the effects of the changes on predicted distributions, usually by comparing with known locations in the native, and perhaps invaded, ranges (Section 5.5.2; Kriticos *et al.*, 2011; Sutherst, 2003; Sutherst & Maywald, 2005). Experimental results or expert knowledge can be used to set parameters; these may require subjective adjustment so that they are directly relevant to the long-term averaged climate data (Section 5.5.4) used in the model. Underpinning the model with as many experimentally derived parameters as possible lowers the reliance on location data and should ultimately produce a more biologically relevant model, provided the experimental data are correct and relevant to field conditions.

Final mapped values include the annual average esoclimatic index (equations 5.1 and 5.2) and annual average growth index. The model is estimated using weekly data so that seasonal variation in suitability can be inferred. This can be a major advantage over applications of correlative models that do not include seasonality predictors. Variation in climatic suitability across years can also be explored through the use of yearly rather than long-term averaged data and based on the assumption that these yearly variations are meaningful to the species. The components of the final indices are multiplicative (equations 5.1 and 5.2), meaning that a low value for any will result in a low prediction. Each component index is scaled from 0 to 1, meaning that each included component contributes equally to the outcome.

The weekly growth index is

$$GI_W = TI_W \times MI_W \times RI_W \times SV_W \times LI_W \times DI_W \,, \tag{5.1}$$

where the indices on the right side are weekly temperature, moisture, radiation, substrate, light and diapause indices, respectively.

The esoclimatic index is

$$EI = GI_A \times SI \times SX \,, \tag{5.2}$$

where $GI_A$ is the annual growth index (mean of $GI_W$), $SI$ is the annual stress index (comprising multiplicative cold, dry, heat and wet stresses) and $SX$ is the annual stress interaction index (comprising multiplicative cold–dry, cold–wet, hot–dry and hot–wet stresses).

Authors refer to this as a process-oriented or mechanistic model (e.g. Kriticos & Leriche, 2010) because (1) the model components consider environmental impacts on the species in a growth and stress framework, similar to process-based population models; and (2) growth and stress are calculated for weekly time steps across the year, mimicking population responses. However, the common use of species data to help fit CLIMEX models creates a clear distinction from the mechanistic models described in Section 5.3.3.

The strengths of CLIMEX for prediction of potential distributions are that it provides a coherent framework for including a range of information (expert knowledge, laboratory data, geographic locations and records of relative abundance) and simple tools for exploring the effect of competitors and mutualists on species distributions. Its authors have emphasised the importance of understanding both the ecology of the species and the frailty of the data, and they have invested time and effort into explaining the model and correcting poor applications. The component indices (e.g. figure above) are restricted to being relatively simple and are constructed so they must define physiological limits, meaning that they should predict sensibly outside their ranges. Nevertheless, if the model is used to predict to novel climates and if species locations are the only available data, the model will be uninformed about the species response in the novel climates, as for other SDMs (Section 5.5.5).

The structure and assumptions of CLIMEX bring limitations for pest risk mapping, as do those of any model. As explained in the Section 5.5, reliance on location data has consequences for the modelled niche (Section 5.5.1) and for sensitivity to sample size (Section 5.5.2). The model structure might be incorrect for some species; responses might be more complex or smoother than the programmed piecewise linear model, and growth and stress might not comprise multiplicative responses to variables that are equally weighted. The model mainly focuses on climate, and inference will be limited (particularly for species with few presence records) if other abiotic variables, biotic interactions, dispersal limitations and disturbances also have an impact on presence records.

While CLIMEX has been widely applied, many modellers choose alternative methods of analysis. Their reasons may include: (a) corporate ownership of CLIMEX influencing cost and willingness of public data modellers to use it; (b) limitation to one software implementation that restricts innovations by users, programmable links to other commonly used software (e.g. R) and use of batch files for sensitivity analyses ; (c) a perception that the coarse gridded output provides less useful spatial detail than that attainable from SDMs applied to finer scale data (this may well be a false impression, depending on the quality of input data, and it is also a historic problem because finer grain data are now available; Kriticos *et al*., 2012); (d) temporal extent: the existing climate data packaged with the program spans from 1961 to 1990 and this may not be relevant to recent invasions; and (e) possibly an aversion to methods that appear to require more research and perhaps more subjective decisions.

## Box 5.3 Overview of modelling methods for correlative species distribution models

A plethora of methods exist for modelling equilibrium species distributions, and a growing body of reviews and texts describes and compares them (Elith & Leathwick, 2009a, 2009b; Franklin, 2010; Guisan & Peterson, 2006; Peterson *et al*., 2011; Renner et al. 2015; Thuiller *et al*., 2008; Zimmermann, 2000). Table 5.1 provides examples of several techniques with key references and invasive species mapping examples. All of the methods have free versions available. Here, I will simply give an overview of the main categories of models and the important differences affecting their use for invasive species modelling.

One set of methods (the true presence-only methods) models environments at presence locations, making no comparison with the range of environments in the broader landscape or at absence sites. Envelope methods are one example. These define the hyper-rectangle that bounds species records in multi-dimensional environmental space, in some cases dealing with relative frequencies of records (e.g. by quantifying percentiles of the distribution). Variables can be weighted equally or unequally, or the response to the most limiting variable can be used for prediction (as in BIOCLIM; Nix, 1986). Related techniques (Franklin, 2010) use distance metrics, such as the Gower metric or Mahalanobis distance, to predict the environmental similarity between records of occurrence and all unvisited sites. A modern machine learning method, the one-class support vector machine, has also been applied to modelling invasive species (Drake & Bossenbroek, 2009; Guo *et al*., 2005). This focuses on finding boundaries that optimally separate occupied environments from all others.

Conceptually, the appeal of this group of methods is that it deals directly with the most common type of data available – presence-only records – and requires none of the additional decisions or assumptions about relevant regions, implied absences *etc.* that discriminative techniques require. This group is dependent on a representative sample of presence locations (as are others), and is adversely affected by bias in the records (e.g. towards urban centres; Aikio *et al*., 2010) because there is generally no information on what has been sampled. PO methods suffer from the problem that they cannot distinguish between landscape availability of environments and habitat suitability, because they include no analysis of available conditions. PO methods are also subject to the usual problems of chance correlations with irrelevant predictors. Some techniques are somewhat biologically unrealistic (e.g. those that equally weight variables). Nevertheless, some are currently methods of choice in biosecurity because they are relatively simple to use and interpret.

All other methods require comparison of presence points with some other class. Some methods were especially developed for modelling equilibrium distributions based on presence-only data (e.g. ENFA, GARP and MaxEnt, Table 5.1). Others are techniques designed for modelling binomial (i.e. two class) data (or in some cases counts or continuous responses) which can be adapted in various ways if used with presence-only species records. Examples include regression and classification methods such as generalised linear models and generalised additive models, decision trees, ensembles of trees including boosted regression trees and random forests. Artificial neural networks are also used. Details of how these methods work are varied and best left to dedicated publications (Table 5.1). All are fitted to species records and environmental data. Many rely on additive terms within the model (e.g. generalised linear models, generalised additive models, boosted regression trees and MaxEnt), which means that even if conditions are suboptimal according to one variable, another can compensate. In contrast, non-parametric multiplicative regression (Table 5.1) is based on multiplicative terms and is therefore more like CLIMEX (Box 5.2) in model structure. Many are capable of modelling interactions between variables (i.e. the response to one variable depends on the value of another). Common applications of several (e.g. generalised linear models and generalised additive models) tend to ignore this capacity, whereas others (e.g. boosted regression trees, random forests and MaxEnt) allow it by default.

Comparisons of methods show that for modelling species at equilibrium, the methods vary in their abilities to retrieve known responses and predict within the training range of the data (Elith & Graham, 2009; Elith *et al*., 2006; Heikennen *et al*., 2007; Moisen & Frescino, 2002). For instance, MaxEnt, tree ensembles and regression methods flexible enough to fit ecologically plausible relationships tend to perform well. Comparisons for invasive species modelling are more difficult because the truth about the potential distribution in the invaded range is unknown. There seems to be a general opinion emerging that smoother models (ones less tightly fitted to the known records) are more likely to predict well, because they do not focus on details of the sampled distribution that might result from survey biases, local responses to biota and so on. Smoother models can be fitted for methods capable of highly complex fits by limiting degrees of freedom and model complexity (e.g. Elith *et al*., 2010; Falk & Mellert, 2011; Merow et al. 2014). I do not think there is enough information yet to make strong conclusions about this idea, although the reasoning seems logical. Studies with artificial species would be useful but are rare.

More generally, in my opinion, a good approach for choosing a particular method is to consider information on its known performance, theoretical aspects of how it works and technical details, including whether its settings can be easily altered and explored and whether it will run well with the types and amounts of data likely to be used. Understanding how a method works, and the implications of default or selected settings, is particularly important for invasive species. Further comments on correlative models, particularly the challenges in using them for pest risk mapping, are included in the discussion of important issues (Sections 5.5.1 to 5.5.7).

## 5.5.2    Issue 2: How species records affect the predicted distribution

All pest risk mapping methods benefit from accurate records across the full native range of the species. This will be universally true because the aim is to characterise all environments in which the species can persist. *Accurate* includes both locational accuracy and taxonomic accuracy. Locational accuracy refers to whether the co-ordinates properly represent the sample to a precision relevant to the grain of the environmental data, while taxonomic accuracy refers to whether the record is truly for the species of interest (Anderson, 2012; Elith & Leathwick, 2009a Elith *et al*., 2013; Funk & Richardson, 2002; Hortal *et al*., 2008; Reddy & Davalos, 2003; Robertson *et al*., 2010; Schulman *et al*., 2007). Record date is also important to accuracy because the record needs to be relevant to the temporal range covered by the available predictors.

Number of records, and their frequency in both environmental and geographic space, has varying importance depending on the modelling method. For instance, CLIMEX can be affected by the number of records, depending on the amount of physiological data available. Without physiological data, CLIMEX requires at least one record in each of the important combinations of environmental conditions (the axes of the environmental space defined by the predictors) inhabited by the species (Lawson *et al*., 2010). Geographic proximity of records is unimportant in CLIMEX, and having more than one record in a given environmental combination does not help model fitting, except to confirm that the conditions are suitable. Having few records most limits the number of parameters that can be meaningfully fitted in CLIMEX when the records are from locations with similar climates. In these cases, some indices have to remain undefined, or a range of values fitted and their effects on the outcome evaluated (van Klinken *et al*., 2009).

Similar limitations apply to correlative SDMs because response data (in this case, species records) are needed to fit model parameters, and having few records limits how many parameters can be fitted, that is, they limit the complexity of the model (in regression, this concept is called events per variable; Harrell Jr, 2006). Further, most correlative SDM methods use the relative frequency of records in different environments to determine relative suitability and sample bias will affect them. This problem is particularly severe for presence-only data (i.e. records of presence that are unaccompanied by records of absence) because there is no information on survey effort, including where the species was not found (Phillips *et al*., 2009). A model may reflect biases in survey effort more than the distribution of the species. There appears to be little research targeted at defining typical biases for invasive species records (e.g. if collectors tend to record presences in unexpected environments rather than randomly), although in the equilibrium SDM literature, research on quantifying biases and methods for dealing with them in models is gradually emerging (e.g. Dorazio, 2014; Fithian *et al.,* 2014; Hortal *et al.*, 2008; Phillips *et al.*, 2009; Warton *et al.,* 2013). There are some examples for invasive species (Wolmarans *et al.*, 2010; Wu *et al.*, 2005), but the topic needs ongoing attention. Even if the samples are a random sample of the species distribution, distance between records should be checked. Correlative SDMs assume that each record is an independent sample, which is untrue for records in very close proximity (Legendre, 1993). Methods for examining spatial autocorrelation in model residuals are useful for diagnosing problems (Bio *et al.*, 2002; Dormann *et al.*, 2007; Rangel *et al.*, 2006). All of these issues imply that data need to be carefully screened before use. This is particularly important when using data from online databases because errors and duplication of records are extremely common (Graham *et al.*, 2004; Robertson *et al.*, 2010).

The type of data (e.g. presence-only, presence-absence or abundance) is also important. Presence-only data are most often used in invasive species SDMs because they are the most common type available and efforts at digitising and correcting them are active and ongoing (Graham *et al.*, 2004; and see sources for data in Woodbury & Weinstein, 2008; Herborg *et al.*, 2009). Rapidly developing technologies offer intriguing possibilities for gathering and storing data (including citizen science projects and the use of mobile phones to capture images and upload data). However, there are many reasons for preferring presence-absence data for correlative modelling because they provide information on what has been surveyed (see Section 5.5.3). Abundance data would be even more useful for invasive species if they indicated the relative fitness of the species across a landscape (e.g. Hooten *et al.*, 2007; Olfert *et al.*, 2006; van Klinken *et al.*, 2009), but only if such relationships were similar in invaded ranges. Several SDM methods can use, or at least be informed by, abundance data. These include CLIMEX and generalised regression methods that can model count data (e.g. Poisson regression; Fithian & Hastie, 2013; Potts & Elith, 2006). For invasive species, presence-absence and abundance data will only be reliable in regions that have been occupied long enough for the species to have had opportunity to persist (and reach stable population states in the case of abundance data) or to die out. Because the aim is to characterise suitable conditions as comprehensively as possible (Section 5.5.1), it is worth gathering all reliable records that are available (i.e. from multiple sources and surveys, but without creating duplicates). Combining data across different surveys does create some difficulties because differing survey efforts will result in differing densities of presence records, but methods are starting to emerge (Fithian & Hastie, 2013; Fithian *et al.*, 2014; Hulme & Weser, 2011).

A final consideration is whether to restrict the model to one based on native range data or include records from the invaded range. The use of presence or abundance records from the invaded range is a two-edged sword. The advantage is that records from the invaded range are likely to expand the representation of environments and biota (Jiménez-Valverde *et al.*, 2011) and can potentially edge the modelled niche towards the fundamental niche. This is the logic in using records from the invaded range in CLIMEX (e.g. van Klinken *et al.*, 2009), and they can also be useful for strict presence-only (one-class) methods (e.g. Booth, 1990), although the lack of equilibrium in the invaded range brings difficulties for interpreting relative frequencies of occurrence in places with active invasion fronts. For two-class methods (Box 5.3), the use of records from the invaded range creates additional conceptual problems in relation to how to set the non-positive case (see Section 5.5.3) and how to make a composite dataset that reflects consistent survey effort. Several studies support the use of some invaded range data (e.g. Broennimann & Guisan, 2008). In the extreme (i.e. the majority of data from invaded ranges) the lack of equilibrium in that the invaded range is certain to cause problems for correlative models unless sophisticated models are used to adjust for variation in propagule pressure and the geographic (spatial) processes of spread (Cook *et al.*, 2007; Elith *et al.*, 2010; Rouget & Richardson, 2003; Williams *et al.*, 2008). All of these problems relating to lack of equilibrium in the invaded range stem from violation of the basic assumption of SDMs (Franklin, 2010), that records are sufficiently well structured to give information on the environments suitable for the species. A species that is spreading will have records that mix environmental preferences with spatial dispersal limitations, and the effects are difficult to untangle.

### 5.5.3 Issue 3: The different views of background records, pseudo-absences and absences

As discussed in Box 5.3, many of the correlative SDM methods applied to presence-only data compare the presence records (the positive case) with another case (note: see Table 5.1 for method abbreviations used hereafter). This approach is used for equilibrium SDMs based on natural history collections (e.g. museums, herbaria, on-line data portals; Graham *et al.*, 2004) and for quantifying resource use by animals within available areas (Manly, 2002). The meaning of the non-positive case varies in subtle but important ways. For some methods and interpretations, non-positive is taken to mean background, landscape or available locations – conditions that can be characterised independently of where the species is present. That interpretation applies to ENFA and MaxEnt and increasing evidence shows it to be the best approach for modelling presence-only data with logistic regression. Presence-background enables a coherent view of how to use regression models for such data (Fithian & Hastie, 2013; Keating & Cherry, 2004; Phillips *et al.*, 2009; Phillips & Elith, 2011; Renner et al. 2015; Ward *et al.*, 2009). So far, most uses of regression (e.g. generalised linear models, generalised additive models and boosted regression trees) with presence-only or background data use naïve models. These do not specifically deal with the problems of presence-only or background data (e.g. that the background points might have a presence at or near them) and do not attempt to model the actual probability of presence because prevalence is unknown (e.g. Elith *et al.*, 2006). While these appear to work reasonably well in some cases, they are not ideal, and current statistical research unifying ideas of density estimation, inhomogeneous Poisson point process models, logistic regression and MaxEnt (Renner et al. 2015) show how to best treat presence-background data in SDMs.

Other viewpoints treat the non-positive case as absence or pseudo-absence. The term *pseudo-absence* is used interchangeably in the literature to refer to either background or implied absence, but here it will mean implied absence. Methods that avoid presence records in sampling pseudo-absences implicitly accept this second view of the data. These include GARP and some uses of regression. For regression, pseudo-absences are placed either anywhere except where presences occur or outside a geographic or environmental buffer around presence records. For instance, Engler *et al.* (2004) used one model to discover areas with low predicted probability of presence and then sampled these to use as pseudo-absences in regression. The species modelling literature (for both equilibrium and invasive species) includes several suggestions about how to establish sensible locations for pseudo-absences or to define reliable absences in the absence of surveyed absences (Le Maitre *et al.*, 2008; Lobo *et al.*, 2010), and new papers with new suggestions keep emerging. However, the background viewpoint requires fewer ad hoc decisions about both position and number of background or pseudo-absence samples, and allows a more rigorous statistical framework (Renner et al. 2015).

Across both of these interpretations, correlative models require decisions about the extent (i.e. the landscape area) to be sampled for background or pseudo-absence points. Users of GARP and MaxEnt have not always understood the importance of this decision, failing to recognise that the model samples the background from any region with data in the gridded predictor variables supplied by the user. So, for instance, if global maps are used without masks for a species whose native range is within South America, the background will be sampled from the whole world. This implies that the species has had the opportunity to reach anywhere and only occurs in South America (Figure 5.2). Unlimited dispersal opportunity is uncommon. Instead, background extent should be restricted to a region that could reasonably be assumed to have been available to the species (Barve *et al.*, 2011; Elith *et al.*, 2011).

True absence data (through comprehensive survey) are relatively rare, but bring several advantages. For instance, absence data provide information on what has been surveyed, and overcome many problems in survey bias. For invasive species modelling, absence data are only likely to be useful in the native range, unless there is clear evidence in the invaded range that the species has had sufficient time and opportunity to spread to, and persist in, surveyed areas, or unless specialised models are used (e.g. Václavík & Meentemeyer, 2009). There has been some discussion of the disadvantages of absence data in the correlative distribution modelling literature, although to my mind, this is overstated. Biotic interactions, dispersal constraints and disturbances affect the distribution of absences (e.g., Jiménez-Valverde *et al.*, 2008), but presence records will be affected similarly, so these impacts should not be used to argue against using absence data (Elith *et al.*, 2011). Presence-absence records are valuable and worth collecting because they remove the need to assume random surveys or deal with survey bias. The important problem with survey-based absence records stems from imperfect detection (i.e. false negative records; Hirzel & Le Lay, 2008; Jiménez-Valverde *et al.*, 2008), but there are now a number of methods available for dealing with imperfect detection in correlative SDMs (e.g. Eraud *et al.*, 2007; Hooten *et al.*, 2007; Wintle *et al.*, 2004). Data need to be used at a grain (spatial resolution) relevant to the species and application, and fine-scale absences may not be informative (e.g. Falk & Mellert, 2011). CLIMEX does not formally use absence data, although information on absence is required or assumed in fitting stress indices (which bound the geographic distribution). In the face of considerable uncertainty about absence, the effect of various assumptions could be explored in sensitivity analyses of the parameters limiting the stress indices.

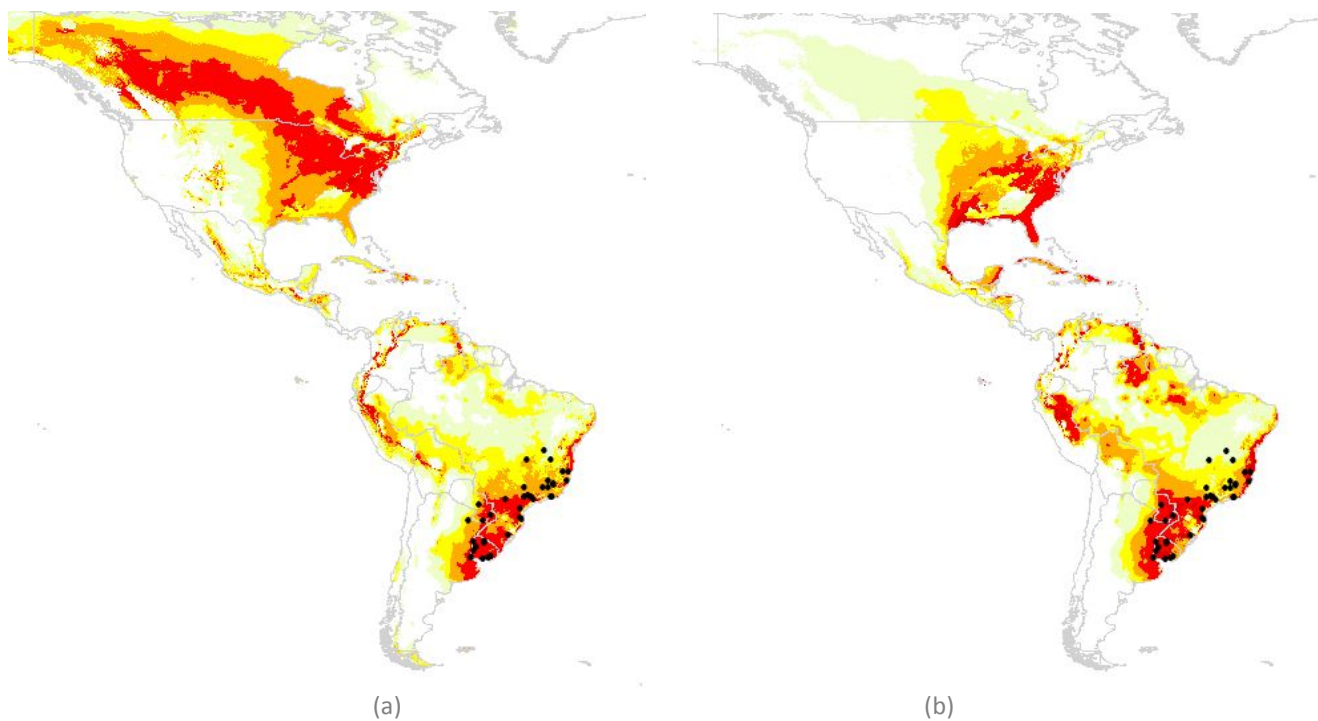

(a)                                                        (b)

**Figure 5.2**: Predictions for the distribution of a hypothetical species located in South America (black dots), using (a) background of South America, and (b) background of the whole world. Modelling method: MaxEnt with linear and quadratic features and five candidate predictors (aridity, humidity, mean temperature of the wettest quarter, highest monthly temperature, minimum monthly precipitation). Colours show the logistic output predictions, red high (0.8 to 1.0) and green low (0.2-0.4). All non-zero predictions are within the environmental range of the training data (i.e. the models are not predicting to novel environments).

What this all means for invasive species modelling is that the user needs to be aware of the assumptions of their method and the requirements for background or absence data. Concepts of the niche and accessible environments are important (Section 5.3). I expect it will take some time to come to a coherent view of the best way to treat these data in correlative methods, so users need to stay abreast of developments.

### 5.5.4    Issue 4: Choice of predictor variables

SDMs for invasive species usually focus on climatic variables. This is partly because climate dominates distributions at the global scale (see discussion of scale in Elith & Leathwick, 2009b) and partly because the only globally coherent terrestrial datasets to date have been climate-based, usually long-term averaged data (for examples and sources see Franklin, 2010; Herborg *et al.*, 2009; Woodbury & Weinstein, 2008). However a broader range of data is becoming available. For terrestrial species, data for soils, topography and measures of climate variability and climate close to the ground are being prepared globally, some at fine resolution (B. McGill & R. Guralnick, personal communication, 2012; Kearney et al. 2014), and coarse resolution marine datasets are now available with a suite of useful predictors (e.g. Tyberghein *et al.*, 2012). Methods are also developed for modelling river networks and summarising environmental conditions throughout the network while taking connectivity into account (Leathwick *et al.*, 2008), although global rivers databases suitable for modelling are currently unavailable. Within the next 10 years, it is reasonable to expect substantial improvements in the quality and quantity of globally complete and biologically relevant predictors for both marine and terrestrial ecosystems. Additional predictors will provide more opportunity to select scales relevant to the modelling problem and use predictors most directly relevant to the species of interest. I expect that predictors that characterise climate extremes and variability and climate close to the ground will be particularly useful for modelling invasive species because they characterise processes and impacts important to species' persistence (e.g. Zimmermann *et al.*, 2009).

This issue of selecting ecologically relevant predictors for correlative models is particularly important for modelling invasive species, and is also discussed in the equilibrium SDM literature. Two viewpoints are evident. The first is that intelligent prior selection of predictors, informed by existing knowledge and theory, will create the firmest foundation for a useful model (Austin & Van Niel, 2011; MacNally, 2000). Mellert *et al.* (2011) call this hypothesis-driven modelling. Austin (2002) argues strongly for the use of proximal predictors that are functionally relevant and best represent the resources and direct gradients that influence species. Distal predictors – such as elevation or ocean depth – rarely affect species distributions directly, but instead do so indirectly through their relationships with proximal predictors such as temperature. The problem with using distal predictors is that they are only relevant to the species through their correlations with proximal predictors, and these correlations tend to change across landscapes and continents. A model fit in one region cannot be guaranteed to predict reliably in another region that has different correlations between variables (Dormann *et al.*, 2013; Elith *et al.*, 2010; Jiménez-Valverde *et al.*, 2011;). The concept of choosing ecologically relevant predictors merges with the thinking behind mechanistic models, and some have discussed the possibility of using mechanistic models to provide physiologically informed predictors for correlative models (Elith *et al.*, 2010; Kearney *et al.*, 2010; Morin & Thuiller, 2009).

The alternative view, that a model should be given the full suite of available predictors so that it can discover the most relevant, is common in data mining and machine learning. Analyses using machine learning methods and hundreds or thousands of predictors have had impressive results in some fields of data analysis, but their success relies on large and unbiased samples of the measured response, and these are rarely available in ecology.

There are many examples of careful selection of variables for invasive species modelling (e.g. Drake & Bossenbroek, 2009; Rodda *et al.*, 2011; Thuiller *et al.*, 2005). It is also not hard to find examples of the alternative approach – the most common being the use of all nineteen temperature and rainfall variables from the Worldclim dataset (Hijmans *et al.*, 2005). So far, there is limited critique in the literature of the effect of these choices, and very few studies include sensitivity analyses of the effect of these choices on model predictions. However, examples are emerging (Le Maitre *et al.*, 2008; Peterson & Nakazawa, 2008; Rodda *et al.*, 2011; Rödder & Lötters, 2010) that confirm the importance of informed selection of directly relevant variables. It is hard to test whether proximal variables can be identified from an available set either by expert knowledge or by modelling, and this needs further exploration. Once a candidate set of variables is selected, iteration between model fitting and evaluation (Sections 5.5.6 and 5.5.7) might suggest the need for changes to the set of candidate variables (e.g. Falk & Mellert, 2011).

Issues of variable selection from extensive geographic information system datasets are not relevant to most CLIMEX analyses (Box 5.2) because the supplied data are limited to a selection of variables available at the time of development and deemed relevant by the authors. These are long-term averaged terrestrial climate data (temperature, rainfall and humidity) that are either site-based (corresponding to ~3 000 meteorological stations worldwide) or gridded at 0.5° (~50 km). Additional data can be added by users, and finer resolution gridded data are now available for use within CLIMEX (Kriticos & Leriche, 2010; Kriticos *et al.*, 2012).

### 5.5.5    Issue 5: Novel environments

In many cases, models fitted to native range data will be predicting into novel environments. This is true for all methods because it is related to the data used to fit the models. The general problem of using correlative models to predict to new geographic regions is often termed *transferability*; when this involves prediction to new environments, *extrapolation* is occurring. Here, the interplay between geographic and environmental space comes to the fore: new geographic regions need not, but often do, harbour new environments.

Protocols have been suggested for dealing with novel environments in CLIMEX. Where predictor values are very different in the invaded range to those for which data are available, it is recommended that parameters for the relevant indices are either not set or a range of likely options examined (van Klinken *et al*., 2009). Much of the early correlative SDM literature on transferability of models either failed to determine whether novel environments occur or used methods for identifying novelty (such as simple data summaries or principal component analyses) that – while useful – weren't spatially mapped (e.g. Randin *et al*., 2006). This makes the results of these studies difficult to interpret. Mapping novel environments (Elith *et al*., 2010; Mesgaran *et al*., 2014; Williams *et al*., 2007) helps interpretation of model output and guides users as to where predictions may be highly unreliable. Novel environments can occur either because the climates in the invaded range are outside the ranges of the training data as assessed on a univariate basis, or they can occur because new combinations emerge, implying changed correlations between variables. If environments are outside the bounds of the data (whether in univariate or multivariate space), knowledge of how the model extrapolates is essential (see column on partial plots in Table 5.1). That is, outside the range of the training data, what trend does the fitted function follow? It is surprising that there has been so little attention to this in the SDM literature for invasive species, although perhaps that reflects the complexity of the topic. Models are usually fitted over multiple predictors, and the only simple way to assess extrapolation is to view partial response plots and the like (i.e. one variable at a time, where the response over the others is held at some constant value; e.g. Figure 5.3, right column). While useful, this approach does not provide a complete picture. For models including interactions (e.g. models based on decision trees, or regression models with interaction terms), understanding how the model predicts in multi-dimensional environmental space is important (Zurell *et al*., 2012).

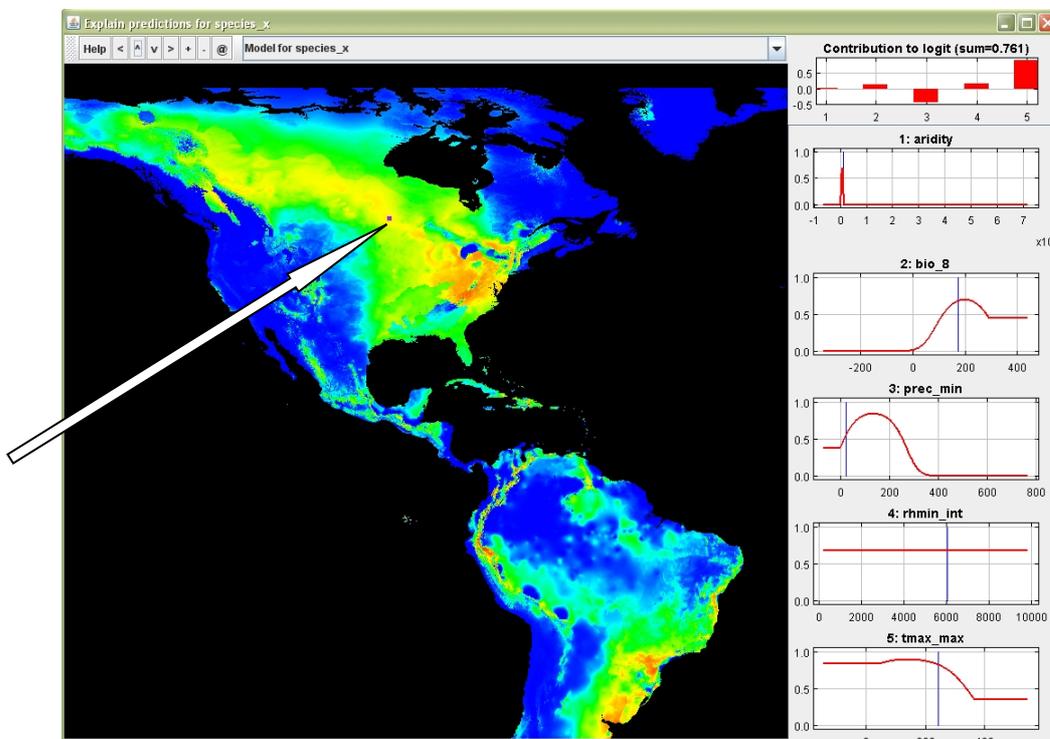

**Figure 5-3** – Example of tool for exploring components of predictions for the species modelled in Figure2. The right pane shows components of the prediction (top panel) and partial plots for each predictor; vertical blue lines in these show the conditions at the location indicated by the arrow. This is from an interactive map produced by MaxEnt (Elith et al. 2010)

The main concern is that using a correlative model to extrapolate beyond the range of the training data is using it outside the realm of safe practice. The models have not been developed for this problem, and methods have not been developed for controlling the models appropriately. Research is only now starting to emerge where models have been carefully controlled through choice of predictors, limiting degrees of freedom in transformations of predictors and controlling the edges of fitted functions (e.g. by weighting data; Mellert *et al*., 2011). I envisage future research on how to fit models that predict well in likely directions of change, how to identify novel environments (including substantially changed correlation structures) and how to control model behaviour to predict in ecologically realistic ways. Simulated data can be useful for exploring how models extrapolate (Fensterer, 2010). Modelling methods that have no facility for visualising fitted functions (Table 5.1) are failing to report vital information, and methods where fitted functions can be controlled (e.g. specialised splines in regression models) will be more easily extended for this application. CLIMEX (Box 5.2) and NAPPFAST (Magarey *et al*., 2007) were specifically developed for invasive species and have functions that are more likely to be appropriately controlled (depending on how well the model is developed). There is no reason why correlative models could not also be developed to use prior information from experts or experiments to control how the model extrapolates.

## 5.5.6    Issue 6: Evaluating predictions

SDMs for species at equilibrium can be evaluated in various ways, for instance, by assessing variable importance and fitted functions and deciding whether the model is consistent with ecological knowledge about the species (Elith & Leathwick, 2009a,b), by exploring the patterns in residuals and by testing predictive performance, ideally at independent sites not used in model training. Emphasis is usually on the last, and statistical summaries including area under the receiver operating characteristic curve, kappa and explained deviance are generally given precedence (Fielding & Bell, 1997; Franklin, 2010; Pearce & Ferrier, 2000).

Some of these methods (particularly the site-based statistical summaries) have been carried over from equilibrium SDM research into invasive species modelling, but they are often not particularly appropriate (Jiménez-Valverde *et al*., 2011). The aim of model evaluation should be to test whether the model is appropriate for its intended application (Rykiel, Jr, 1996). Because prediction in the native range is not the aim, the fact that a model can do this successfully is reassuring but not ultimately a strong test. The problem is clear: the potential distribution in the invaded range is unknown and test data are not available. The main question is whether the model fitted in the native range is relevant to the invaded range. Distributional data in the invaded range are unlikely to provide a reliable test of model performance because the species is likely to be invading; presences may not indicate persistence and absences will be unreliable. More attention should be given to the problem of evaluation, including how to simulate data that is useful for model testing (Austin *et al*., 2006; Fensterer, 2010). Models need to be assessed for their ecological relevance: by using expert knowledge, by sourcing additional data including physiological information or by comparison with completely independent models that do not use distributional records. Evaluation could also address questions about the sensitivity of the model to choices made in the modelling process (see Section 5.5.7). Methods for perturbing or resampling data that tested model behaviour in environments most common in the invaded range might be also useful. Because the problem of predicting potential invasive distribution is – from a modelling viewpoint – quite similar to the problem of predicting changes in distribution with climate change, progress on evaluation methods in that arena is likely to be transferable to invasive species (for an interesting example, see Falk & Mellert, 2011).

## 5.5.7    Issue 7: Dealing with uncertainty

This section relies on a mix of information from equilibrium SDMs and invasive species applications (including models of spread in invaded ranges) because most pest risk mapping examples focus on only one component of uncertainty. Uncertainty in predictions emanates from multiple sources, including those discussed in Sections 5.5.2 to 5.5.4, and choice of modelling method and its settings (Box 5.3, Table 5.1).

While there have been a number of theoretical treatments and reviews of sources of uncertainty in correlative equilibrium SDMs and related fields (Ascough *et al*., 2008; Barry & Elith, 2006; Elith *et al*., 2002; Kangas & Kangas, 2004; Leyk *et al*., 2005; Rocchini *et al*., 2011), relatively little has been done in practice to characterise the effect of likely uncertainties on modelled predictions (but see Dormann *et al*., 2008; Elith *et al*., 2013; Gutzwiller & Barrow, Jr, 2001; Johnson & Gillingham, 2008; Leung *et al*., 2012; van Niel & Austin, 2007). This is largely because it is difficult to quantify errors, and the problem seems overwhelming once possible errors are scoped. Uncertainty is only partly characterised by confidence intervals from models (Elith *et al*., 2002; Kuhn *et al*., 2006). Rocchini *et al*. (2011) emphasise the need for maps of ignorance to depict areas where the reliability of predictions is either known or unknown and suggest potential approaches for producing these.

Most research has targeted important components of uncertainty, including bias in species records (e.g. Argaez *et al*., 2005; Hortal *et al*., 2008; Rodda *et al*., 2011), uncertainty in predictors (Kriticos & Leriche, 2010; van Niel & Austin, 2007), differences between modelling methods (Pearson *et al*., 2006) and different parameterisations of one model (Hartley *et al*., 2006). Ensembles of correlative methods are favoured by some modellers (e.g. Araujo *et al*., 2005; Caphina & Anastácio, 2010; Roura-Pascual *et al*., 2009, Stohlgren *et al*., 2010; Thuiller, 2003) as a means of dealing with the sometimes extreme variation in predictions across methods. Their aim is to emphasise agreement of predictions and to quantify model-based uncertainty. However, these are not problem-free, particularly for invasive species. Ensemble SDM methods are usually based on standard application of the component modelling methods (e.g. generalised linear models, generalised additive models, Mahalanobis distance and boosted regression trees; Table 5.1) with default settings chosen by the ensemble programmer and any weighting of the ensemble components based on predictive performance to some set of sites. Because point-based predictive performance is usually impossible to evaluate meaningfully for invasive species, the ensemble components are often simply averaged (Araújo & New, 2007). It is unclear whether variation between components of the ensemble (i.e. between individual methods) is largely due to unrealistic models that have not been thoroughly explored and evaluated rather than real uncertainty between predictions. In my opinion, use of ensembles is only a good idea if the component models have been rigorously evaluated (e.g. Falk & Mellert, 2011). There are several reasons for this. Available species data sets are rarely so large and error-free that a model can be left to sort out the mess. The shapes of modelled responses require evaluation. Default settings may not be appropriate; the model might be too complex (as is often the case with machine learning methods using standard settings) or too simple (linear fits in GLMs). Extent of extrapolation needs to be evaluated, especially as it interacts with the shape of the modelled response (Section 5.5.6).

A useful approach for exploring uncertainty in any model is to fit multiple parameterisations to test the many judgments made in fitting the model (Elith *et al*., 2013; Ray & Burgman, 2006; Taylor & Kumar, 2012; van Klinken *et al*., 2009). Another angle for exploring uncertainty is to ask what type and amount of uncertainty would lead to a changed decision based on the model, or whether a decision or action is robust to estimated uncertainty (e.g. Elith *et al*., 2013; Moilanen *et al*., 2006; Yemshanov *et al*., 2010; see Chapters xx in this volume). Alternatively, adaptive surveillance approaches can be used by starting with models based on existing information (even if inadequate) and then iteratively updating the models with new information resulting from actions aimed at achieving some mix of management and data collection (McCarthy & Parris, 2008; Rout *et al.*, 2014).

While it might be easier to believe that a model is accurate, it is important to face the range of likely uncertainties and to communicate them in a way that aids decision-making and future data collection. Further research – focusing on how to make practically useful evaluations of uncertainty – will progress informed use of predictions (Venette *et al*., 2010).

## 5.6    Conclusions

Many practitioners will need to use models based on data from the realised niche, whether as a stop-gap measure before better methods are available or because these might remain one of the only options for many species. An obvious question is which method to adopt. In my opinion, because these models require understanding, a better question is what expertise to develop. A skilled analyst is important for understanding the issues; they can also learn more than one method and choose methods that suit their data and species. Methods such as CLIMEX have been specifically developed for invasive species and have some features that make them safer to use (e.g. the way their indices can be controlled to extrapolate beyond the realised niche). These methods will not suit all species and all situations, and it is useful to continue development of other methods and tools. Some researchers are optimistic that correlative models will predict with high precision (e.g. Peterson, 2003); while that may be true for some species at some scales of evaluation, I believe that the issues discussed in this chapter make substantial errors reasonably likely. I am hopeful that ongoing developments will produce models better suited to the task and tools to help practitioners to better understand predictions and their uncertainties.

## 5.7 Acknowledgements

Thanks to Terry Walshe, Yvonne Buckley, Matt Hill and Karl Mellert for thoughtful comments on drafts.